\documentclass{article}
\usepackage[pdftex]{graphicx}
\usepackage{amsmath}

\usepackage{changepage} 
\usepackage[utf8]{inputenc}
\usepackage[backend=biber,style=ieee,]{biblatex}
\title{References}
\usepackage[T1]{fontenc}
\usepackage{subfiles}
\usepackage{url}
\usepackage{physics}
\usepackage{braket}
\usepackage[official]{eurosym}
\usepackage{geometry}
\usepackage{authblk}
\usepackage{caption}
\usepackage{hyperref}
\usepackage{subcaption}
\usepackage{verbatim}
\usepackage[table,xcdraw]{xcolor}
\usepackage{enumitem, kantlipsum}
\usepackage{fancyvrb}
\usepackage{longtable}
\usepackage{lscape}
\usepackage{amssymb,amsmath}
\usepackage{graphicx}
\usepackage{float}
\usepackage{multirow}
\usepackage{tabularx}
\usepackage{marginnote}
\usepackage{amsmath}
\usepackage[nottoc]{tocbibind}
\usepackage{flexisym}
\usepackage{amsmath}
\usepackage{array}
\usepackage{graphicx} 
\usepackage{multirow}
\usepackage{hhline,booktabs,amsmath}
\usepackage{makecell}
\usepackage{xparse}
\usepackage{listings}
\usepackage{tikz}

\title{\textbf{"Toward" Metal-Organic Framework Design by Quantum Computing}}

\author[1,2]{Kourosh Sayar Dogahe}
\author[2]{Tamara Sarac}
\author[1,2]{Delphine De Smedt}
\author[1,2]{Koen Bertels}

\affil[1]{Department of Computer Science and Engineering, Ghent University, Ghent, Belgium}
\affil[2]{QBee B.V. Quantum Accelerated, Leuven, Belgium}
\date{}

\begin{document}
\maketitle

\begin{center}
    \textbf{Abstract}
\end{center}

\begin{adjustwidth}{+1cm}{1cm} 
The article summarizes the study performed in the context of the Deloitte Quantum Climate Challenge in 2023 
\footnote {Deloitte hosts an annual Quantum Climate Challenge aiming to encourage climate-relevant collaborations between sustainability and quantum computing experts. The QBee team won first place in the 2023 challenge among 118 registrations from 33 countries. We want to express our gratitude to Deloitte and IBM Quantum companies for their assistance in providing the necessary materials. Link:  [ \href{https://www2.deloitte.com/de/de/pages/innovation/contents/deloitte-quantum-climate-challenge.html}{Delloite Challenge '23}].} \footnote {We would also like to thank QBee Company for their generous support and for helping us with our Quantum Computing research and development needs. To learn more about QBee Company, please visit their website at: [ \href{https://qbee.eu/}{QBee.eu}].}. We present a hybrid quantum-classical method for calculating Potential Energy Surface scans, which are essential for designing Metal-Organic Frameworks for Direct Air Capture applications. The primary objective of this challenge was to highlight the potential advantages of employing quantum computing. To evaluate the performance of the model, we conducted total energy calculations using various computing frameworks and methods. The results demonstrate, at a small scale, the potential advantage of quantum computing-based models. We aimed to define relevant classical computing model references for method benchmarking.  The most important benefits of using the PISQ approach for hybrid quantum-classical computational model development and assessment are demonstrated.
\vspace{10pt} 


\end{adjustwidth}

\section{Introduction}
With the anticipated progress in quantum technology, quantum computing could bring about significant advantages over classical computing in the future. These include the ability to tackle previously intractable problems and the potential for substantial computational speedup. Computational chemistry is a domain anticipated as among the first ones to potentially benefit from quantum computing. It tackles the behavior of electrons and atoms at the quantum level, and classical computers struggle to accurately simulate those systems. Therefore, it is crucial to develop quantum technology-based computational methods, also in the domain of computational chemistry. 

Quantum computing is a multidisciplinary domain, that combines hardware and software development. Extensive work on quantum computing in the last decade led to the establishment of approaches like the NISQ (Noisy Intermediate-Scale Quantum), which refers to a category of near-term quantum computers handling noisy qubits, resulting from different noise sources such as decoherence, gate errors, and measurement errors \cite{Preskill2018quantumcomputingin}. Going further, there is the PISQ (Perfect Intermediate-Scale Quantum) approach that enables various scientific and industrial communities to step into the quantum computing field by letting the hardware concerns aside and focusing on the quantum computing logic for their specific expert domains \cite{bertels2022quantum}.  The "perfect qubit" is a theoretical concept, defined after the famous talk by Richard Feynman \cite{feynman2018simulating}, that is based on perfect qubit behavior simulation on a classical computer.  Perfect qubits are often used to explore the full potential of quantum computing algorithms and protocols. PISQ and NISQ represent complementary strategies that are expected to converge to an operational field within the next 10-15 years. \cite{Preskill2018quantumcomputingin, bertels2022quantum}


In the framework of Deloitte's Quantum Climate Challenge 2023, an interesting use case was proposed. The primary goal was to investigate how quantum computers may help to improve materials used in Direct Air Capture (DAC) of $\mathrm{CO_2}$.  We approached the challenge from the chemistry domain following the PISQ approach, with a goal to assess the structure of metal-organic frameworks (MOFs), with particular attention to MOF74 \cite{Dzubak2012}.  

The main observation indicates that a potential advantage of quantum computing-based models over classical methods can be demonstrated at a small scale. Namely, quantum computation based results reach the accuracy of the methods that incorporate an analytical approach, going beyond the accuracy obtained by conventional classical methods employment. Our attempt to scale up the quantum-based computation revealed some drawbacks of present quantum models and tools.

 The paper is structured as follows. We first present the goal of the challenge and our approach to it in the Problem Definition Section. In the Method section, we provide detailed descriptions of all the methods and implementations. This is followed by a Results and Discussion section where the methods are benchmarked and evaluated. The article concludes by summarizing our impressions of the quantum model's application and announcing future steps.

\section{Problem definition}

MOFs consist of two key components: an inorganic metal cluster and an organic molecule. The choice of metal and organic components influences the structure and, consequently, the properties of the MOF. There is a broad range of possibilities for combining metallic cations and organic ligands to construct the MOF74 \cite{Mitra2022}. A well-designed MOF is capable of efficiently capturing $\mathrm{CO_2}$ in DAC filters while hindering the clogging of the filter by other gases. Screening for potentially suitable MOF74 structures involves calculating the Dissociation Energy (DE) of a complex formed by MOF74 and gas molecules. To calculate the DE of a MOF-gas complex, we require Potential Energy Surface (PES) scans of both the complex and individual molecules, as depicted in Equation 1.

\begin{center}
\begin{equation} \label{eq:1} 
    \mathrm{DE = PES_{gas-ion} - (PES_{gas} + PES_{ion})}
\end{equation}
\end{center}

 The method for computing PES scans was developed using  Variational Quantum Eigensolver (VQE), which is defined as a hybrid quantum-classical, variation principle-based algorithm\cite{peruzzo_2014}. The model is exemplified using the interaction between a magnesium ion ($\mathrm{Mg^{2+}}$) and carbon dioxide gas ($\mathrm{CO_2}$). 

Due to the structural complexity of the MOF-gas system, the latter must be down-scaled for the analysis of quantum-based results. Fig. 1 presents an example of system simplification, from a network of MOF unit cells, passing through a singular unit cell with six open reaction sites (metallic ions), and culminating in one preferential binding site (metal ion) - gas interaction, neglecting the organic part \cite{Dzubak2012}.
 
\begin{figure}[H]
  \centering
  \includegraphics[width=0.8\textwidth]{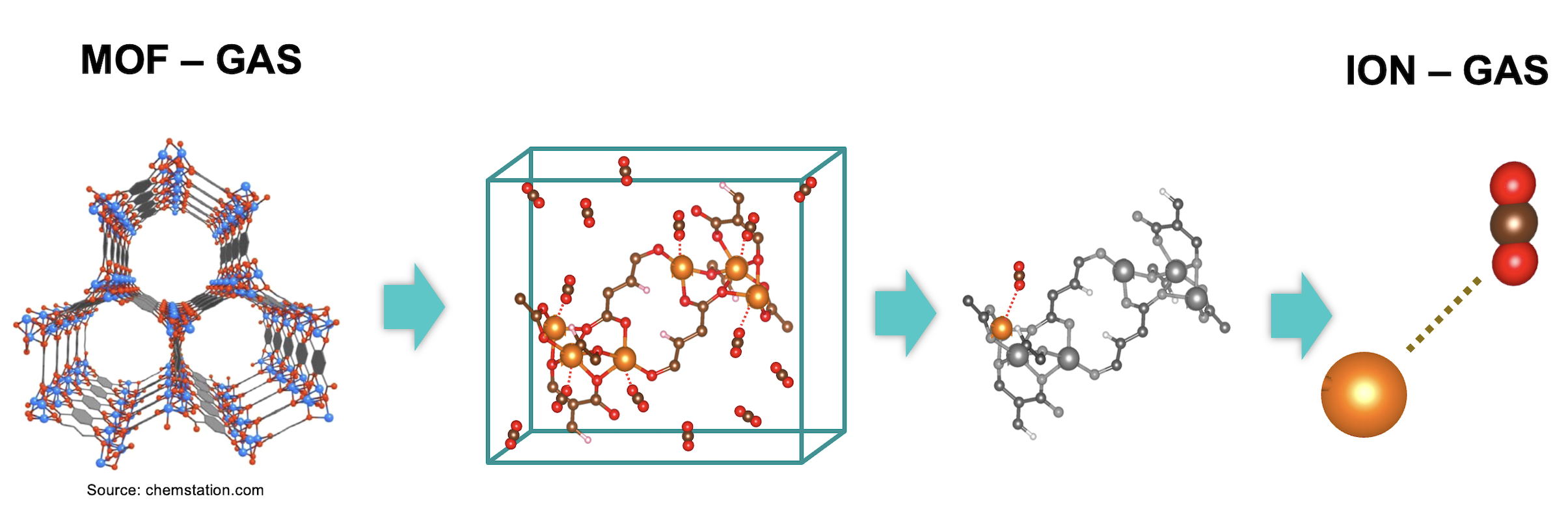}\label{fig:f234}
  \caption{The MOF structure simplification scheme
 }
\end{figure}

 The electronic structure of $\mathrm{CO_2 - Mg^{2+}}$, modeled using the minimal basis set STO-3G, encompasses a total of 48 spin orbitals and approximately 700,000 fermionic terms. During the construction of the Electronic Hamiltonian, these spin orbitals require encoding using roughly 48 qubits, leading to the generation of over 12 million Pauli terms. To this day, quantum computing resources are constrained by a limited number of qubits and low circuit depth for both physical quantum computers and classical emulation of quantum computing processes. Consequently, the number of spin orbitals was reduced within the scope of this study using the AS Transformation, as described in the Method section.

 The challenge was to demonstrate the potential benefits of a quantum computing-based method. To evaluate this, we benchmarked total energy calculations using various computing approaches and methods, including: 
\begin{enumerate}[label=(\alph*)]
    \item VQE algorithm as a hybrid quantum-classical computation implemented by perfect qubits; 
    \item VQE algorithm as a hybrid quantum-classical computation implemented by superconducting physical qubits, i.e., using quantum hardware;
    \item \textit{Ab-initio} classical methods: the low-cost Restricted Hartree-Fock (RHF) and the higher-cost Coupled Cluster single and Double excitation (CCSD) \cite{RevModPhys.79.291};
    \item Complete Active Space (AS) Configuration Interaction (CASCI) analysis as the classical reference method \cite{ROOS1980157}.
\end{enumerate}

A 3D potential energy surface scan in quantum mechanics offers a comprehensive exploration of molecular behavior by systematically altering atomic positions and calculating corresponding potential energies \cite{Wang2019-kb, Yan2014-hc}. Such scans are valuable for interpreting chemical reactions, reaction pathways, and mechanisms, as well as pinpointing transition states and energy barriers.  The potential of VQE to produce PES scans that consider multiple degrees of freedom has been also examined through the creation of $\mathrm{CO_2 - Mg^{2+}}$ 3D PES scan. 

Finally, a draft approach to compute the energy of a full unit cell was proposed. It is based on what is known as the classical hybrid approach, involving the combination of low-cost computational methods on a larger scale; and local high-cost computational methods focused on the open reaction sites of a molecule \cite{Valenzano2011}.  Following a similar logic, we applied a low-cost classical method on the unit cell - CO2 interaction and quantum computing-based calculations on open reaction sites, which are metal ions in this case. The obtained results are well-converged and they follow the expected trend.

\section{Method}

\textbf{Geometry Optimization.} The Avogadro software was used to define the optimized geometry of $\mathrm{CO_2 - Mg^{2+}}$. To evaluate different degrees of freedom within the molecule, we selected the Z-matrix coordinate system. In the optimization process using the Universal Force Field (UFF) method, we executed the Steepest Descent method with 500 steps.

\textbf{Active Space Transformation.} The AS Transformation involves the construction of an AS configuration by selecting a certain number of orbitals from the full electronic structure. It is challenging to find the AS configuration that contributes the most to the total energy since it relies on varying factors such as the Highest Occupied Molecular Orbitals (HOMOs) and Lowest Unoccupied Molecular Orbitals (LUMOs) within the electronic structure \cite{ROOS1980157}. Quantum computation is performed only for the defined AS. The number of HOMO and LUMO orbitals chosen to construct the AS configuration was varied. By reducing the number of contributor spin orbitals in the model, the computational resources required for the calculations were also reduced (see Table 1).

\begin{table}[H]
\caption{The overview of resource requirements for $\mathrm{CO_2-Mg^{2+}}$ at different levels of approximation}
\scriptsize \centering

\begin{tabular}{|c|c|c|c|c|c|}
\hline
\textbf{\begin{tabular}[c]{@{}c@{}}System \\ Size\end{tabular}} & \textbf{\begin{tabular}[c]{@{}c@{}}System\\ Abv.\end{tabular}} & \textbf{\begin{tabular}[c]{@{}c@{}}No. Spin Orbitals \\ = No.  qubits\end{tabular}} & \textbf{\begin{tabular}[c]{@{}c@{}}No. Fermionic\\ Terms\end{tabular}} & \textbf{\begin{tabular}[c]{@{}c@{}}No. Pauli\\ Terms\end{tabular}} & \textbf{\begin{tabular}[c]{@{}c@{}}Hamiltonian's\\ Circuit Depth\end{tabular}} \\ \hline
Full                                                            & Full                                                           & 48                                                                                  & 730612                                                                 & 12207792                                                           & 254329                                                                         \\ \hline
5 HOMO - 5 LUMO                                                 & 5h5l                                                           & 20                                                                                  & 20136                                                                  & 142940                                                             & 7147                                                                           \\ \hline
4 HOMO - 4 LUMO                                                 & 4h4l                                                            & 16                                                                                  & 8292                                                                   & 46608                                                              & 2913                                                                           \\ \hline
3 HOMO - 3 LUMO                                                 & 3h3l                                                            & 12                                                                                  & 2664                                                                   & 11076                                                              & 923                                                                            \\ \hline
2 HOMO - 2 LUMO                                                 & 2h2l                                                            & 8                                                                                   & 564                                                                    & 1544                                                               & 193                                                                            \\ \hline
1 HOMO - 1 LUMO                                                 & 1h1l                                                            & 4                                                                                   & 36                                                                     & 60                                                                 & 15                                                                             \\ \hline
\end{tabular}
\end{table}

 The following procedure was used to select the orbitals: Molecular Orbitals (MOs) index was generated using the PySCF programming tool \cite{pyscf} and the electronic Hamiltonian corresponding to the desired AS configuration was constructed using the AS transformation module.

 \textbf{Energy Computation on Hybrid Quantum-Classical Framework. }The PES computation was performed using the IBM Qiskit Nature Module \cite{Qiskit} with a PySCF interface. We employed the VQE algorithm and designed it using the Qiskit library.  The VQE algorithm was executed on the \textit{statevector} backend using perfect qubits but also on quantum hardware using the \textit{Ibmq--nairobi} backend  via superconducting qubits. Henceforth, we will refer to simulations based on perfect qubits as "VQE - perfect q." and simulations based on physical qubits as "VQE - physical q."

 The singlet Hartree-Fock state was defined as the initial state for all quantum computing-based computations. The variational form was constructed by the Unitary Coupled Cluster Single and Double excitation (UCCSD) using fully-entangled hardware-efficient ansatz. The ansatz implemented on the \textit{ibmq--nairobi} backend (VQE-physical q.) was parameterized by the rotational parameters obtained from VQE-perfect q. computations. The fermionic terms of selected AS configurations (1h1l, 2h2l, 3h3l, and 4h4l; see Table 1) were mapped to a qubit Hamiltonian in the minimal basis set (STO-3G) using the Bravy--Kitaev transformation. The two-qubit reduction technique was employed to decrease the number of qubits by two in each generated qubit Hamiltonian circuit. The SLSQP and SPSA optimizers were used to optimize the ansatz parameters toward minimizing the electronic structure energy in \textit{statevector} and \textit{ibmq--nairobi} simulators, respectively. To mitigate errors incurred during qubit readout in physical q.-based implementation, we applied readout error mitigation at\textit{ resilience level one}.

 \textbf{Energy Computation on Classical Framework.} The PES of the full $\mathrm{CO_2 - Mg^{2+}}$ electronic structure was calculated in PySCF, applying RHF and CCSD methods on the STO-3G basis set to obtain the full electronic structure energy references. To obtain a relevant total energy reference, with respect to the quantum computing-based method and AS Transformation applied, the localized Full Configuration Interaction analysis was performed.  This was done by exact {Diagonalization} of different AS configuration matrices embedded into the Hartree-Fock spin orbitals, which is known as CASCI. The diagonalization was done using \textit{NumpyEigensolver}.

\section{Results and Discussion}
The choice of an energy calculation method in quantum chemistry depends on the balance between computational resources and the desired level of accuracy. The system state is defined by the Schrödinger equation and several methods are employed to solve it, varying in accuracy level. The mean-field approach, such as HF, provides a simplified but computationally efficient representation of electron-electron interactions, considering electrons as independent in an average field generated by all other electrons. Since it neglects the electrons correlation effect,  it leads to a low accuracy calculation, especially for strongly correlated systems \cite{szabo2012modern}. CCSD is an extension or correction with respect to the HF. It accounts for some electron correlation, regarding various levels of excitation and often yields highly accurate results for many molecular properties. The computational cost for CCSD is expensive with respect to the other classical methods \cite{shavitt2009many}.  The most accurate manner to simulate the  molecular systems is by considering all possible electronic configurations and interactions,  this could be done by the FCI approach. FCI becomes computationally intractable for large systems due to its exponential scaling. Local application of FCI on a subset of important configurations (CASCI), strikes a balance between accuracy and computational cost \cite{ROOS1980157}. The configuration interaction size selected is dictated by the computational power of the machine used and even the most powerful classical machines fail to perform FCI on large molecular systems.

VQE belongs to the category of variational methods, unlike FCI which is a deterministic method.  VQE aims to compute an  upper bound for the lowest possible expectation value of an observable with respect to a trial wave function. The objective of the VQE is therefore to find a parametrization of the trial wave function, such that the expectation value of the Hamiltonian (energy equation) is minimized. VQE is assumed to achieve higher accuracy than classical mean-field methods but at a lower cost compared to FCI. However, it is worth mentioning that its accuracy depends on factors like the choice of initial state, ansatz, optimization method, execution approach, etc. \cite{mcardle2020quantum, TILLY20221}.

\textbf{The Total Energies at the Optimized Geometry.} To compare the accuracy level of energy computation for the optimal inter-nuclear distance, the total energies of  the electronic structure of $\mathrm{CO_2 - Mg^{2+}}$   (ground state energy) are shown as a function of AS configuration size (number of HOMOs and LUMOs orbitals involved), see Fig. 2.  Fig. 2 (A) schematically presents the different AS configurations. The HOMO and LUMO orbitals chosen to create an AS configuration are marked blue. The remaining inactive orbitals are marked red. Different markings highlight that CASCI and VQE computation were applied on AS configurations (blue orbitals), while the mean-field (HF) was applied on inactive space (red). To obtain the total energy,  CASCI/ VQE was embedded into the mean field calculation.


\captionsetup[figure]{skip=10pt} 

\begin{figure}[H]
  \centering
   \captionsetup{justification=centering} 

  {\includegraphics[width=0.6\textwidth]{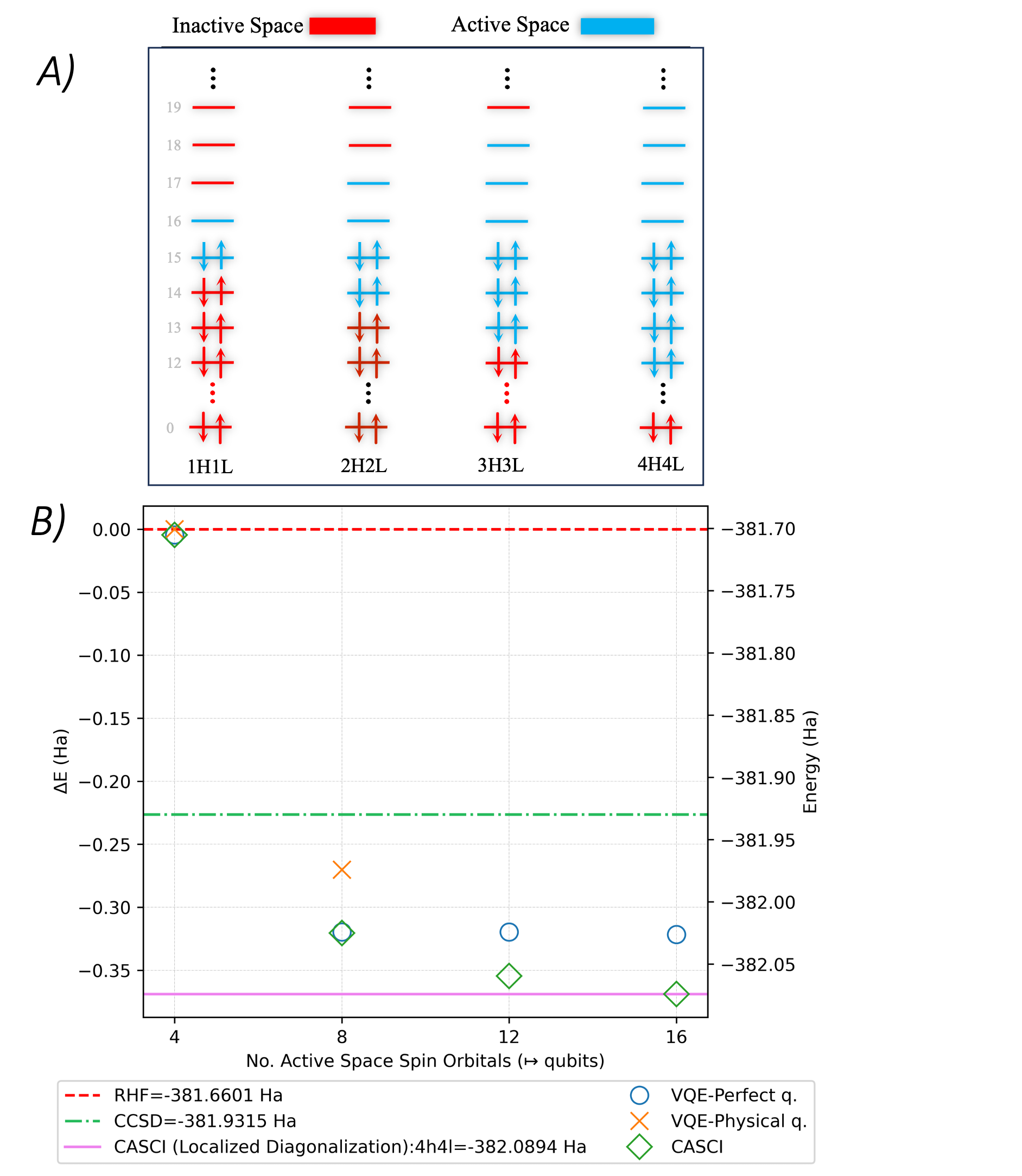}\label{fig:f12}}
  \hfill
  \caption {A) The schematic presentation of AS configurations B) The $\mathrm{CO_2 - Mg^{2+}}$ relative energy with respect to the RHF($\Delta E$) and absolute energy (Energy) at optimal bond length. For the methods where AS transformation was applied, $\Delta E $ and Energy are presented as a function of AS Configuration interaction size.}
\end{figure}

 In Fig. 2 (B), RHF and CCSD results shown as dotted lines, illustrate classical method thresholds of low and higher accuracy, respectively. One should note that these two methods were applied to the entire $\mathrm{CO_2 - Mg^{2+}}$ electronic structure. Two different energy scales were used to evaluate the data: total energy (right axes) and total energy improvement  with respect to the RHF accuracy level (left axes, $\Delta$E = $\mathrm{E-E_{RHF}}$). The lowest computed energy, alongside its corresponding methods and available computing resources, is used to represent the highest accuracy level (solid line).
 
 It can be observed that accuracy significantly increases when more than 1h1l orbitals are involved. AS configuration size is expected to influence energy value \cite{Veryazov2016}. VQE-Perfect q. 2h2l result  is comparable to CASCI 2h2l. This demonstrates that the hybrid model performs comparably well as its classical reference, and both are above the CCSD level. The deviation observed for VQE-physical q. 2h2l implementation in comparison to the perfect qubit and classical outcome is a result of erroneous hardware execution. With increasing AS space size, considering the investigated range, the quantum computing-based method shows no significant improvement in the accuracy level. Contrarily, energy computed with CASCI decreases with increasing AS size. The lowest energy was achieved for CASCI implementation with 4h4l. 

\textbf{Potential Energy Surface Scan.} Fig. 3 (A-D)  presents total energy evolution as a function of inter-nuclear distance i.e. PES scan, obtained by various methods (see section Method) for 1h1l, 2h2l, 3h3l, and 4h4l configurations.  RHF and CCSD are also shown in the figure representing full electronic system classical computations.


\begin{figure}[H]
  \centering
  \captionsetup{justification=centering}
  {\includegraphics[width=0.9\textwidth]{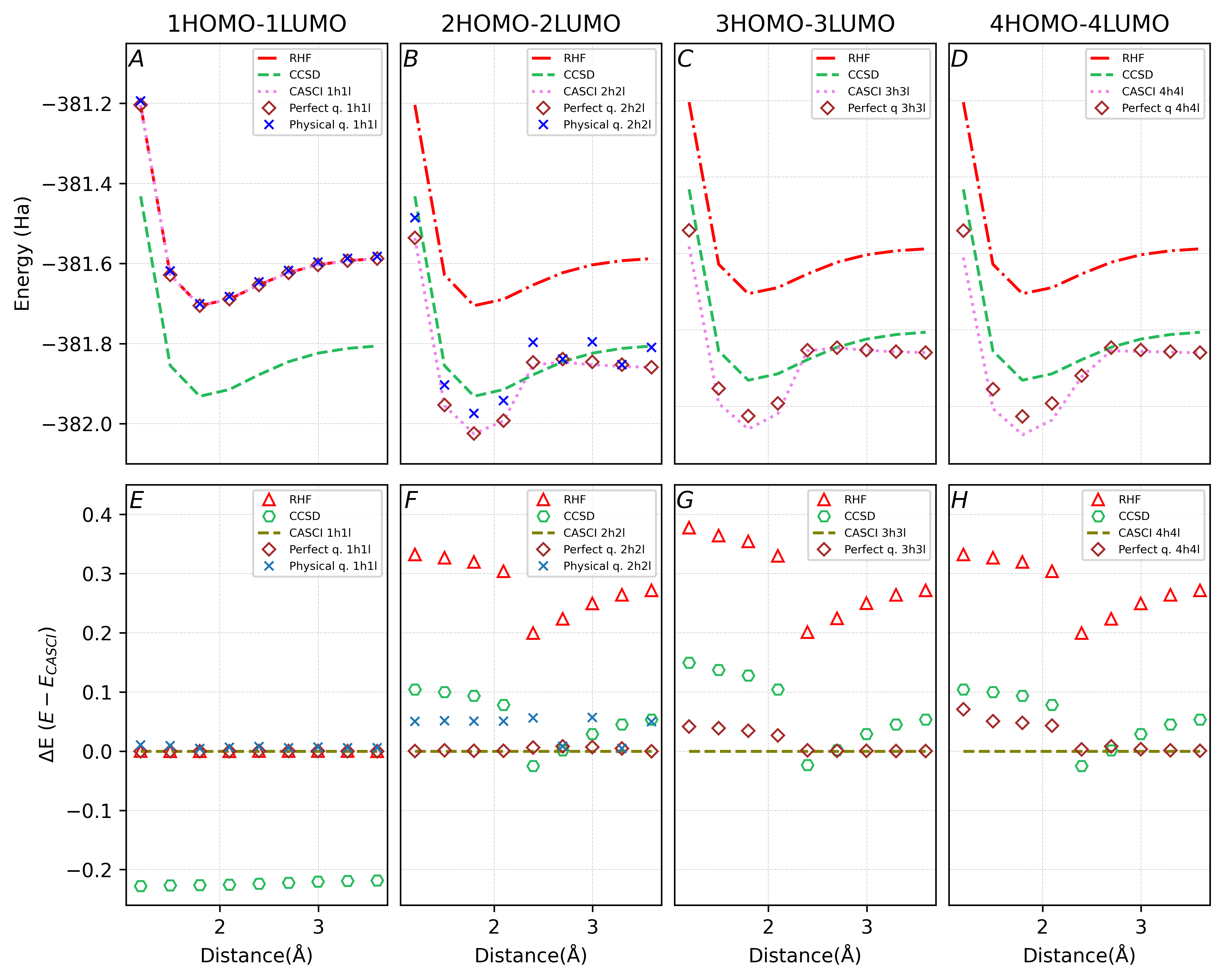}\label{fig.f11}}
  \hfill
  \caption{ (A) (B), (C), and (D) are the $\mathrm{CO_2 - Mg^{2+}}$ PES scans for 1h1l, 2h2l, 3h3l, and 4h4l AS configuration as a function of internuclear distance, respectively, including the RHF and CCSD results for the full electronic structure.  (E), (F), (G), and (H) represent the deviation VQE results with 1h1l, 2h2l, 3h3l, and 4h4l configurations from their own corresponding Localized CASCI energy reference, respectively. RHF and CCSD deviations are related to the calculation of the full electronic structure including the RHF and CCSD results.}
\end{figure}

For 1h1l computation, VQE and CASCI computations meet RHF accuracy level. This can be expected, due to a low number of orbitals considered within 1h1l AS configuration. In general, one can notice that as the AS size expands, accuracy gradually increases. However, this trend does not hold true for the VQE calculations, with both perfect q. and physical q. Nonetheless, the data obtained by the VQE algorithm are well-converged and demonstrate a similar trend to the energy evolution calculated on classical computers. 2h2l, 3h3l, and 4h4l PES scan trends show a difference with respect to the full configuration PES scans (RHF and CCSD)  at about 2.6 \AA. Namely, AS results exhibit energy  decay for distances larger than $\sim$ 2.6 \AA.

 The observed deviation between VQE and CASCI results is analyzed in Fig 3 (E-H). For a particular AS-Configuration, CASCI computation is taken as a reference, and other computations are illustrated relative to their corresponding Localized CASCI. For the 1h1l configuration, it can be observed again that all the results overlapped with CASCI, except the CCSD, which reaches higher accuracy. VQE-perfect q. 2h2l exhibits a reasonable overlap with CASCI 2h2l, both being more accurate with respect to the RHF and CCSD.  VQE-physical q. deviation resulting from erroneous behavior of actual quantum hardware can be observed, as expected. However, 3h3l and 4h4l configurations suffer from discrepancies between VQE and CASCI data at low distances.  It is expected that these deviations become more obvious with the AS size increasing. Still, it is interesting to notice that energies obtained by VQE, for both perfect q. and physical q, seem to follow their corresponding CASCI results. Moreover, for the STO-3G minimal basis set, these surpass the accuracy of the CCSD, despite CCSD holding a reputation of a computationally demanding approach. 

 The potential of quantum computing-based methods can be envisioned through the fair agreement between the observed VQE-perfect q. and CASCI results. To further improve energy computation, one should enlarge an AS size. However, this is computationally expensive, and classical resources would be consumed before reaching a significant number of orbitals, especially for large and strongly correlated systems. With the assumption that quantum computing resources will grow, it can be presumed that quantum computing-based methods play a key role in reaching better accuracy levels for chemical system energy calculations. It's important to emphasize that the performance of quantum hardware would need to improve concurrently, as indicated by the deviation in the results between VQE-physical q. and VQE-perfect q.  It was observed that the discrepancy between VQE and CASCI increases with increasing the AS configuration size, which suggests that VQE performance decreases with enlarging an electronic structure matrix. This demonstrates that further algorithm improvements and development are needed in order to reach an industry-relevant quantum computing application.



\textbf{3D Potential Energy Surface Scan.} Different adsorption orientation angles for the $\mathrm{CO_2-Mg^{2+}}$ interaction were incorporated with existing calculations over the bond distance. This resulted in a 3D PES scan (Fig. 4), which is obtained by VQE-Perfect q.

\begin{figure}[H]
  \centering
  \captionsetup{justification=centering} 
  \subfloat[]
  {\includegraphics[width=0.45\textwidth]{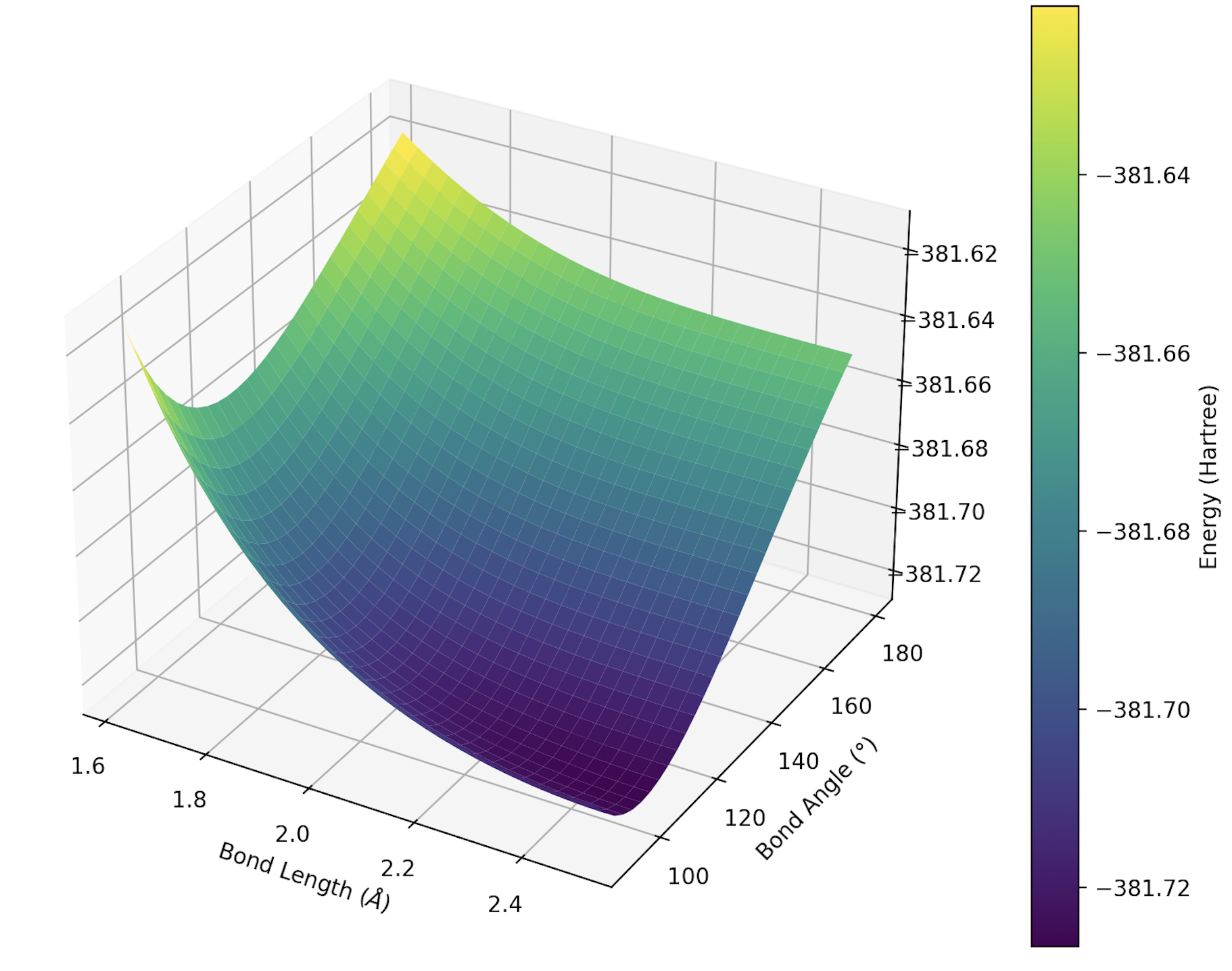}\label{fig:f121}}
  \hfill
  \subfloat[ ]{\includegraphics[width=0.45\textwidth]{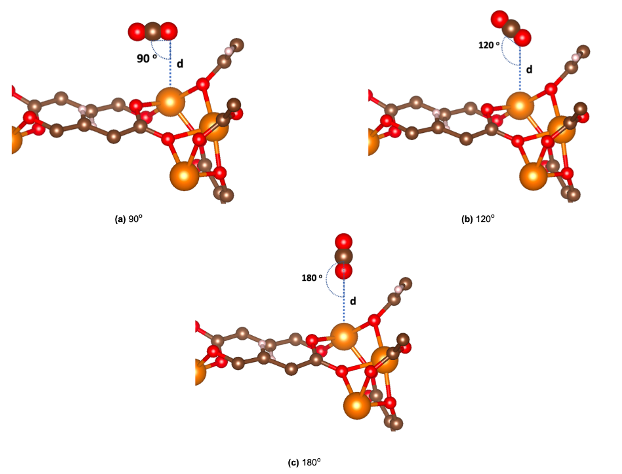}\label{fig:f162}}
  \caption{(a) 3D PES scan of the $\mathrm{CO_2 - Mg^{2+}}$ obtained by variation of two degrees of freedom and, (b) Schematic of different orientations of $\mathrm{CO_2}$ on a hypothetical $\mathrm{CO_2 - Mg^{2+}}$ embedded in a scaled-up MOF74 structure}
\end{figure}

Results presented in Fig 4. show that VQE can be used successfully to manipulate two degrees of freedom. Still, one should note that the calculation is not complete, with respect to the range of analysis needed for forming chemistry-relevant conclusions about, for example MOF - CO2 reaction path.

\textbf{The MOF74 Unit Cell Energy Computation.} The idea behind the proposed computation is to draft a direction toward the MOF74 full unit-cell energy calculation. A local energy correction (LEC) method was applied as schematically presented in Fig 5.  The energy of the MOF74 unit cell, without $\mathrm{CO_2}$ gases involved $(\mathrm{E(RHF_{S})})$ was calculated with RHF, representing the low-cost method. The energy of the $\mathrm{CO_2}$ interaction with the most active molecule part, which is six open reaction sites (6 $\mathrm{Mg^{2+}}$ ), was calculated with VQE-perfect q. $(\mathrm{E(VQE_{S})})$. In addition,  RHF was applied to the open reaction sides $(\mathrm{E(RHF)_{C})})$, as it is required for final energy determination. The final $\mathrm{E_{LEC}}$ was computed as presented in Eq. 2. Existence of preferential binding sites and entanglement between chemical structures calculated by RHF and VQE were not considered within this study. 

\begin{center}
\begin{equation} \label{eq:2} 
    \mathrm{E_{LEC} = E(RHF_{S}) + E(VQE_{C}) - E(RHF_{C})}
\end{equation}
\end{center}

\begin{figure}[H]
  \centering
  \captionsetup{justification=centering} 
  {\includegraphics[width=0.65\textwidth]{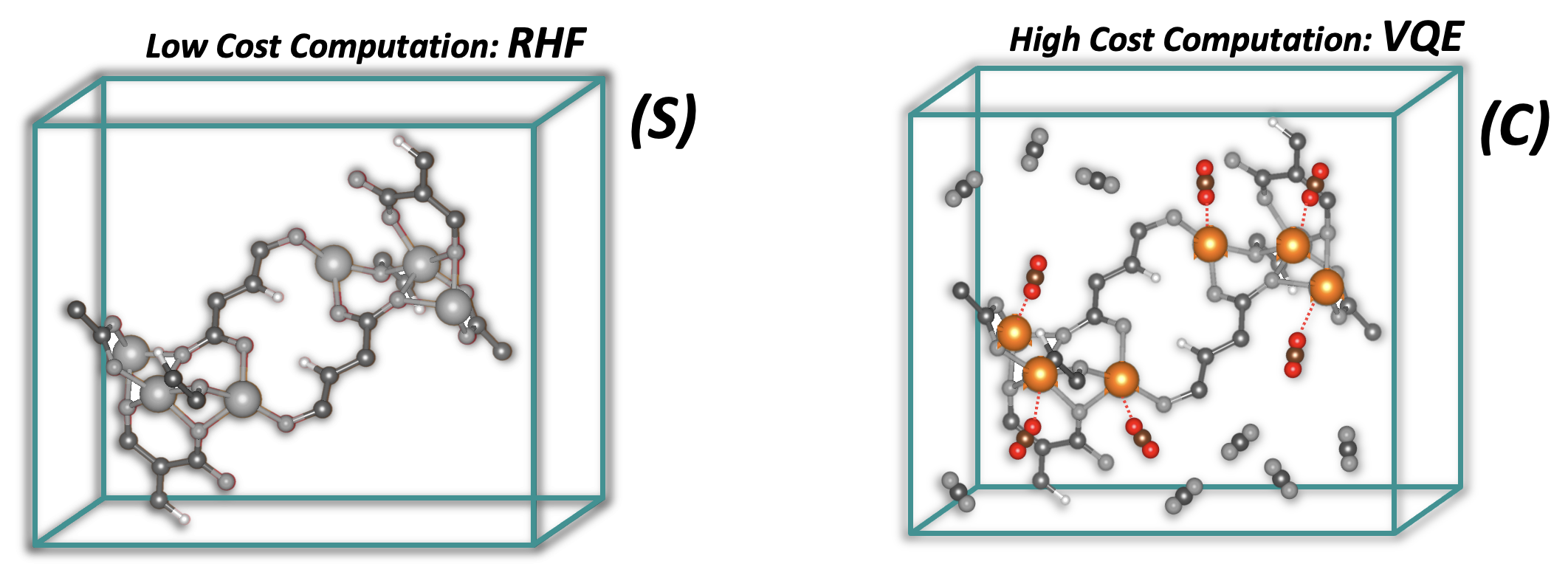}\label{fig:f7}}
  \caption{Schematic of LEC method application for the PES computation of 6Mg-MOF74 - $\mathrm{CO_2}$ complex}
\end{figure}

\begin{figure}[H]
  \centering
  \includegraphics[width=0.5\textwidth]{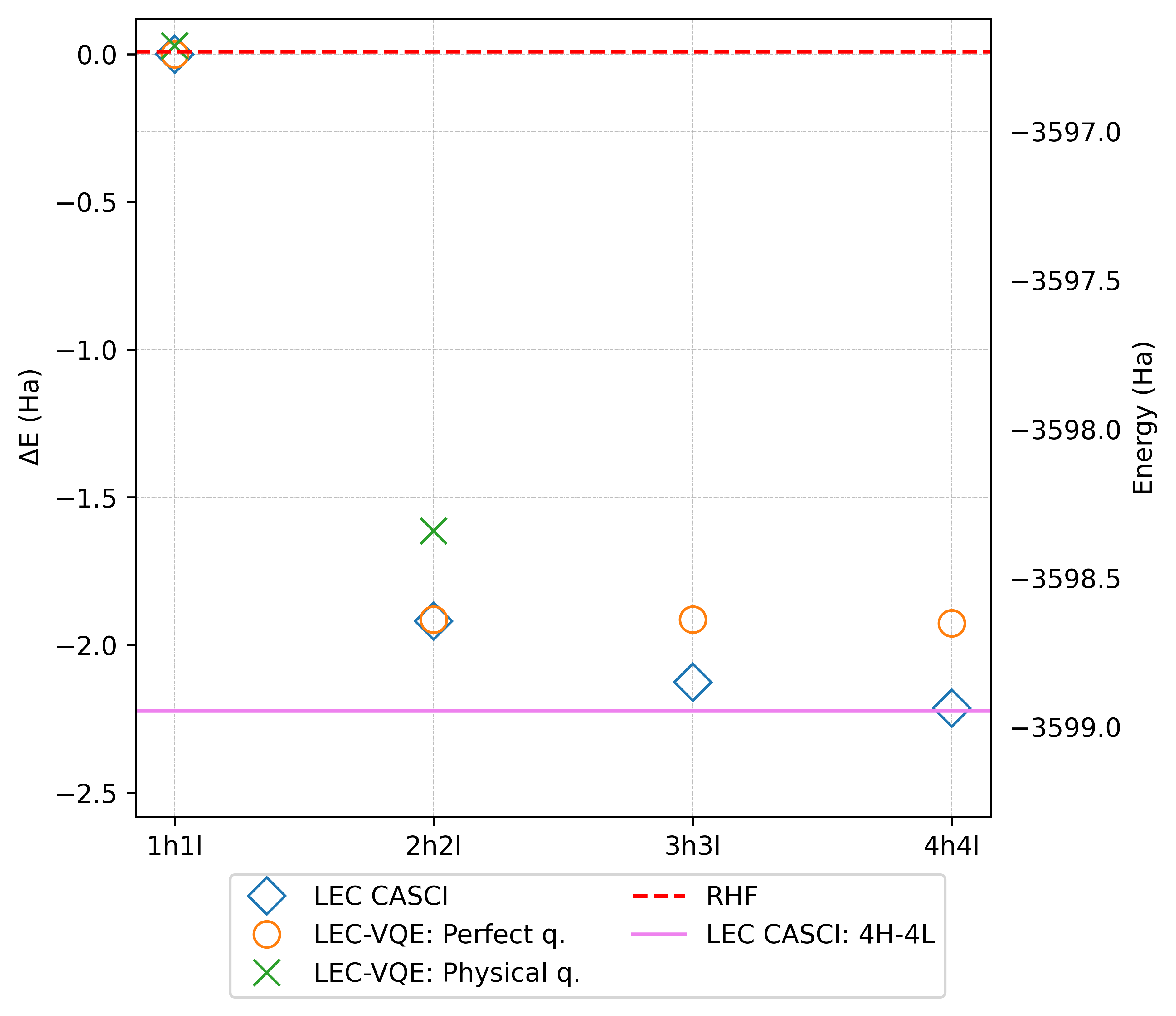}\label{fig:f10}
  \caption{(\textit{Left Axis}:) The $\mathrm{CO_2 - 6Mg-MOF74}$ relative total energies with respect to the RHF ($\Delta$E) and \textit{(Right Axis}:) absolute total energies at optimal bond length. For the methods where CASCI and VQE approaches were applied, both are presented as a function of growing AS configurations.
 }
\end{figure}

 Fig. 6 presents the total energy computation for CO2 - 6Mg MOF74 interaction and relative energy with respect to the RHF reference value, $\Delta$E. In addition,  VQE-perfect q. 4h4l result is presented as a solid line to illustrate thresholds in the same way as in Fig 2. It can be observed that the RHF accuracy level is improved by applying quantum computing-based local energy correction. 
\section{Conclusion and future work}

This work aimed to evaluate if MOF design could be enhanced by quantum computing in the future. To investigate that, a quantum computing-based energy calculation model was constructed following the PISQ approach and a comparison of perfect and physical q.-based implementation was done in parallel with \textit{ab-initio} (RHF, CCSD) and CASCI computations. The model was based on complete active space configuration interaction. The potential of quantum has been demonstrated through well agreement of VQE-perfect q. results with localized FCI-based calculations (CASCI). However, an attempt to scale up the model to involve more molecular orbitals and hardware implementation revealed drawbacks of current quantum computing: low number of qubits available, erroneous behavior of qubits, and algorithm design challenges. A pathway toward MOF 74 full unit cell energy computation was proposed. It is based on low and high-cost model applications on different MOF unit cell building structures. Well-converged results followed the trend as expected.  However, the energy computation needs to involve existing structure interactions in a more representative manner.  Perfect qubit-based simulation proved to be a valuable approach to perform model evaluation. In addition, VQE-perfect q. results were used to parameterize VQE-physical q. implementation, which shortens hardware runtimes.  Perfect qubit simulation was used to draft a pathway toward modeling 3D and full unit cell PES scans. These procedures do not reach chemistry-relevant outcomes yet, and implementing them on hardware brings no advantage. However, it is needed to validate the proof of concept and the PISQ approach is suitable. These indicate some of the most important benefits of the PISQ approach.  The authors plan to expand the work by performing hardware implementation beyond the 2h2l AS configuration, to enable further analysis of the quantum potential. More investigation will be done to advance unit cell energy computation.

\printbibliography

@book{shavitt2009many,
  title={Many-body methods in chemistry and physics: MBPT and coupled-cluster theory},
  author={Shavitt, Isaiah and Bartlett, Rodney J},
  year={2009},
  publisher={Cambridge university press}
}

@article{pyscf,
author = {Sun, Qiming and Berkelbach, Timothy C. and Blunt, Nick S. and Booth, George H. and Guo, Sheng and Li, Zhendong and Liu, Junzi and McClain, James D. and Sayfutyarova, Elvira R. and Sharma, Sandeep and Wouters, Sebastian and Chan, Garnet Kin-Lic},
title = {PySCF: the Python-based simulations of chemistry framework},
journal = {WIREs Computational Molecular Science},
volume = {8},
number = {1},
pages = {e1340},
doi = {https://doi.org/10.1002/wcms.1340},
year = {2018}
}

@article{peruzzo_2014,
  title={A variational eigenvalue solver on a photonic quantum processor},
  author={Peruzzo, Alberto and McClean, Jarrod and Shadbolt, Peter and Yung, Man-Hong and Zhou, Xiao-Qi and Love, Peter J and Aspuru-Guzik, Al{\'a}n and O’brien, Jeremy L},
  journal={Nature communications},
  volume={5},
  number={1},
  pages={1--7},
  year={2014},
  publisher={Nature Publishing Group}
}

@misc{Qiskit,
    author = {{Qiskit contributors}},
    title = {Qiskit: An Open-source Framework for Quantum Computing},
    year = {2023},
    doi = {10.5281/zenodo.2573505}
}

@article{Veryazov2016,
author = {Veryazov, Valera and Malmqvist, Per Åke and Roos, Björn O.},
title = {How to select active space for multiconfigurational quantum chemistry?},
journal = {International Journal of Quantum Chemistry},
volume = {111},
number = {13},
pages = {3329-3338},
doi = {https://doi.org/10.1002/qua.23068},
url = {https://onlinelibrary.wiley.com/doi/abs/10.1002/qua.23068},
eprint = {https://onlinelibrary.wiley.com/doi/pdf/10.1002/qua.23068},
year = {2011}
}

@article{ROOS1980157,
title = {A complete active space SCF method (CASSCF) using a density matrix formulated super-CI approach},
journal = {Chemical Physics},
volume = {48},
number = {2},
pages = {157-173},
year = {1980},
issn = {0301-0104},
doi = {https://doi.org/10.1016/0301-0104(80)80045-0},
url = {https://www.sciencedirect.com/science/article/pii/0301010480800450},
author = {Björn O. Roos and Peter R. Taylor and Per E.M. Sigbahn}
}

@ARTICLE{Wang2019-kb,
  title    = "Accurate Potential Energy Surfaces for the Three Lowest
              Electronic States of {N((2)D}) + {H(2)(X(1)$\sum$(g}) (+))
              Scattering Reaction",
  author   = "Wang, Dequan and Shi, Guang and Fu, Liwei and Yin, Ruilin and Ji,
              Youbo",
  journal  = "ACS Omega",
  volume   =  4,
  number   =  7,
  pages    = "12167--12174",
  month    =  jul,
  year     =  2019,
  address  = "United States",
  language = "en"
}

@ARTICLE{Yan2014-hc,
  title    = "Potential energy surfaces and reaction pathways for
              light-mediated self-organization of metal nanoparticle clusters",
  author   = "Yan, Zijie and Gray, Stephen K and Scherer, Norbert F",
  journal  = "Nature Communications",
  volume   =  5,
  number   =  1,
  pages    = "3751",
  month    =  may,
  year     =  2014
}

@article{Preskill2018quantumcomputingin,
  doi = {10.22331/q-2018-08-06-79},
  url = {https://doi.org/10.22331/q-2018-08-06-79},
  title = {Quantum {C}omputing in the {NISQ} era and beyond},
  author = {Preskill, John},
  journal = {{Quantum}},
  issn = {2521-327X},
  publisher = {{Verein zur F{\"{o}}rderung des Open Access Publizierens in den Quantenwissenschaften}},
  volume = {2},
  pages = {79},
  month = aug,
  year = {2018}
}

@Article{Dzubak2012,
author={Dzubak, Allison L.
and Lin, Li-Chiang
and Kim, Jihan
and Swisher, Joseph A.
and Poloni, Roberta
and Maximoff, Sergey N.
and Smit, Berend
and Gagliardi, Laura},
title={Ab initio carbon capture in open-site metal--organic frameworks},
journal={Nature Chemistry},
year={2012},
day={01},
volume={4},
number={10},
pages={810-816},
issn={1755-4349},
doi={10.1038/nchem.1432},
url={https://doi.org/10.1038/nchem.1432}
}

@misc{bertels2022quantum,
      title={Quantum Computing -- from NISQ to PISQ}, 
      author={Koen Bertels and Tamara Sarac and Aritra Sarkar and Imran Ashraf},
      year={2022},
      eprint={2106.11840},
      archivePrefix={arXiv},
      primaryClass={quant-ph}
}

@Article{Mitra2022,
author={Mitra, Abhishek
and Hermes, Matthew R.
and Cho, Minsik
and Agarawal, Valay
and Gagliardi, Laura},
title={Periodic Density Matrix Embedding for CO Adsorption on the MgO(001) Surface},
journal={The Journal of Physical Chemistry Letters},
year={2022},
day={18},
publisher={American Chemical Society},
volume={13},
number={32},
pages={7483-7489},
doi={10.1021/acs.jpclett.2c01915},
url={https://doi.org/10.1021/acs.jpclett.2c01915}
}

@article{RevModPhys.79.291,
  title = {Coupled-cluster theory in quantum chemistry},
  author = {Bartlett, Rodney J. and Musia\l{}, Monika},
  journal = {Rev. Mod. Phys.},
  volume = {79},
  issue = {1},
  pages = {291--352},
  numpages = {0},
  year = {2007},
  publisher = {American Physical Society},
  doi = {10.1103/RevModPhys.79.291},
  url = {https://link.aps.org/doi/10.1103/RevModPhys.79.291}
}

@Article{Valenzano2011,
author={Valenzano, Loredana
and Civalleri, Bartolomeo
and Sillar, Kaido
and Sauer, Joachim},
title={Heats of Adsorption of CO and CO2 in Metal--Organic Frameworks: Quantum Mechanical Study of CPO-27-M (M = Mg, Ni, Zn)},
journal={The Journal of Physical Chemistry C},
year={2011},
day={10},
publisher={American Chemical Society},
volume={115},
number={44},
pages={21777-21784},
issn={1932-7447},
doi={10.1021/jp205869k},
url={https://doi.org/10.1021/jp205869k}
}

@book{szabo2012modern,
  title={Modern quantum chemistry: introduction to advanced electronic structure theory},
  author={Szabo, Attila and Ostlund, Neil S},
  year={2012},
  publisher={Courier Corporation}
}

@article{mcardle2020quantum,
  title={Quantum computational chemistry},
  author={McArdle, Sam and Endo, Suguru and Aspuru-Guzik, Al{\'a}n and Benjamin, Simon C and Yuan, Xiao},
  journal={Reviews of Modern Physics},
  volume={92},
  number={1},
  pages={015003},
  year={2020},
  publisher={APS}
}

@article{feynman2018simulating,
  title={Simulating physics with computers},
  author={Feynman, Richard P and others},
  journal={Int. j. Theor. phys},
  volume={21},
  number={6/7},
  year={2018}
}

@article{TILLY20221,
title = {The Variational Quantum Eigensolver: A review of methods and best practices},
journal = {Physics Reports},
volume = {986},
pages = {1-128},
year = {2022},
note = {The Variational Quantum Eigensolver: a review of methods and best practices},
issn = {0370-1573},
doi = {https://doi.org/10.1016/j.physrep.2022.08.003},
url = {https://www.sciencedirect.com/science/article/pii/S0370157322003118},
author = {Jules Tilly and Hongxiang Chen and Shuxiang Cao and Dario Picozzi and Kanav Setia and Ying Li and Edward Grant and Leonard Wossnig and Ivan Rungger and George H. Booth and Jonathan Tennyson}
}

\end{document}